\documentstyle[graphicx,float]{article}

\begin{document}
\title{Entanglement, identity and disentanglement in two-atom spontaneous emission}
\date{}
\author{Pedro Sancho \\ Centro de L\'aseres Pulsados CLPU \\ Parque Cient\'{\i}fico, 37085 Villamayor, Salamanca, Spain}
\maketitle
\begin{abstract}
Two recent experiments studying the potential effect of entanglement
on the emission properties of excited atoms produced in molecular
photodissociation have been interpreted in conflicting ways. We
present a theoretical analysis of the problem, showing that the
experimental results can be explained by a combination of three
processes: entanglement, exchange effects associated with the
identical nature of the atoms and disentanglement by spontaneous
emission. According to our approach these experiments provide the first
verification of the phenomenon of disentanglement by spontaneous
emission.
\end{abstract}


\section{Introduction}

Two recent experiments have explored the potential role of
entanglement in the emission properties of pairs of excited hydrogen
atoms produced in the photodissociation of $H_2$ molecules. In the
original experiment \cite{jap}, the coincidence time spectra of the
emitted photons were measured. The authors concluded that the decay
time constant of the spectra was dependent on the degree of entanglement
in the system. Later \cite{com}, it was signaled that the
above argument cannot be considered conclusive because the method of
evaluation of the decay constant is based on the independence of the
two emission processes, a controversial assumption for entangled
systems. In that paper it was also suggested that the correct
description should be based on a method of photon detection able to
distinguish the temporal ordering of the two emission processes. A
setup fulfilling that condition has been presented in \cite{bel};
where the emission and detection times can be measured, determining
the temporal ordering of the two events. Those authors concluded,
in marked contrast with \cite{jap}, that the emission effects are
not dependent on entanglement.

We present in this paper a theoretical analysis of the problem aimed
at settling the question. During the analysis we realize that
entanglement effects are actually present in the problem, but they
alone cannot explain the complete phenomenon. Two other ingredients,
exchange effects and disentanglement by spontaneous emission, must
enter into the physical picture. The argument supporting this
conclusion is simple: the compatibility between the emission
distributions with and without temporal ordering imposes a precise
relation between the emission rates of the pair of atoms and that of
a single isolated atom. That relation leads, via some simple
calculations, to the conclusion that the first photon emission
depends on entanglement and identity and the second on the process
of disentanglement by spontaneous emission \cite{ye,ey}.

In order to clarify the problem further from the conceptual point of
view we discuss, following well-known arguments in
entanglement theory, how some measurable properties of the two-atom
system depend on the non-separable character of the initial state
and others do not. To be concrete, the temporally ordered distributions
are dependent, whereas those without temporal ordering and the coincidence time spectra are not.

The phenomenon of disentanglement by spontaneous emission
\cite{ye,ey} enters into the physical picture of the problem
providing an elegant explanation of the behavior of the second
photon emission. Conversely, setups of the type of \cite{jap} and \cite{bel}
experimentally verify the phenomenon. This is a relevant result
because it is its first direct demonstration. Disentanglement
by spontaneous emission is a particular instance of the more general
process of disentanglement induced by the environment (sometimes
taking place in a finite time, then being denoted entanglement
sudden death). Environment-induced disentanglement was first
experimentally tested in \cite{bra} using an optical arrangement.
Reference to later tests and developments can be found in \cite{ey}.
However, to our knowledge, the loss of entanglement associated
with spontaneous emission has not been experimentally verified.

\section{Photon emission distributions}

In the experiments in \cite{jap} and \cite{bel} $H_2$ molecules are
photodissociated producing pairs of hydrogen atoms in excited
states. Two detectors are placed at fixed directions outside the
cell containing the molecules, and the arrival times at both
positions are registered. They are denoted $t_1$ and $t_2$
(events registered at detectors $1$ and $2$). In \cite{jap} only the
distribution $t_1 - t_2$ is measured. In \cite{bel} the synchrotron
clock signal is also recorded. With this additional record one can also
determine the times of the first ($t_f$) and second ($t_s$)
photon detections.

We denote by $N_f(t)$ and $N_s(t)$ the number of first and second
emitted photons recorded between the photodissocation ($t=0$) and
$t$. Consequently, the total number of emitted photons registered in
that time interval is
\begin{equation}
N(t)=N_f(t)+N_s(t)
\end{equation}
If we denote by $n_0$ the number of photodissociated molecules at
the initial time the numbers of first and second photons are \cite{com}
\begin{equation}
N_f(t)=n_0 (1-e^{-\Gamma _f t})
\end{equation}
and
\begin{equation}
N_s(t)=N_f(t)- n_0 \frac{\Gamma _f}{\Gamma _s - \Gamma _f} (e^{-\Gamma _f t} - e^{-\Gamma _s t})
\end{equation}
with $\Gamma _f$ and $\Gamma _s$ the first and second emission rates.

The total number can also be expressed in terms of the number of
photons registered at each detector, $ N(t)=N_1(t)+N_2(t)$. Note
that we have $N_1(t)=N_2(t)$ because each emission process takes
place in a random direction and for a large number of repetitions of
the experiment both distributions become statistically similar. This
property has been experimentally verified in \cite{bel}. This result
can also be written as $N_i(t)= N(t)/2$ with $i=1,2$.

The experiment in \cite{bel} shows that the distributions $N_i$ have
the form $N_i= n_0 (1- e^{-\Gamma  t})$ and, consequently
\begin{equation}
N(t)=2 n_0 (1- e^{-\Gamma t})
\end{equation}
with $\Gamma $ the emission rate for a single atom or, equivalently,
of atoms in product states.

It is immediately seen that the three distributions are compatible
when the conditions
\begin{equation}
\Gamma _f= 2\Gamma
\label{eq:ff}
\end{equation}
and
\begin{equation}
\Gamma _s =\Gamma
\label{eq:ss}
\end{equation}
hold. Naturally, this is the value obtained by the fits of the first
and second emission curves \cite{urb}.

We represent the three detection patterns graphically. We have done
the previous presentation in terms of the total number of photons of
a type detected between the initial time and $t$. It can also be
done, as in \cite{bel}, using the distributions of photons of each
type detected at a given time $t$. For instance, for the first
photons the number of detections in a short interval $\Delta t$
around time $t$ is $ N_f(t+\Delta t)-N_f(t) \approx n_0 \Gamma _f
e^{-\Gamma _f t}\Delta t $. Then the normalized number of detections
per unit time at time $t$ is $n_f(t)=( N_f(t+\Delta t)-N_f(t))/n_0
\Delta t \approx \Gamma _f e^{-\Gamma _f t}$. We represent the three
distributions in Fig. 1. The detection patterns coincide with the
count patterns in \cite{bel}.

We devote the rest of the paper to understanding the physical meaning
of the two conditions (\ref{eq:ff}) and (\ref{eq:ss}).

\begin{figure}[H]
\center
\includegraphics[width=8cm,height=7cm]{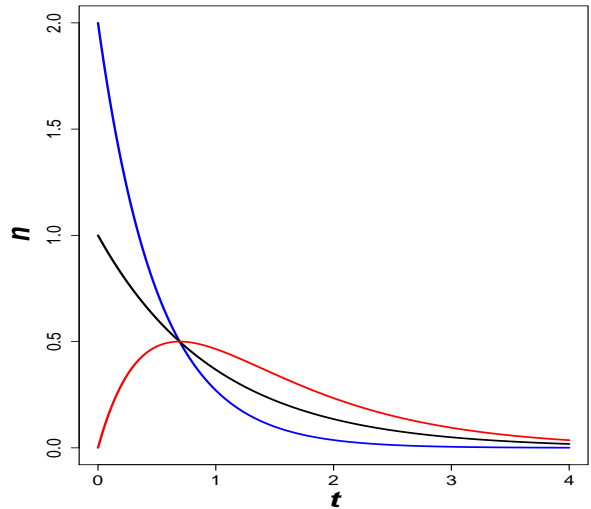}
\caption{Temporal dependence of the normalized number of detections
per unit time of photons at each detector ($n_i, i=1,2$, black
line), first photons ($n_f$, blue line), and second photons ($n_s$,
red line). The time and detection numbers are represented in units
of $\Gamma ^{-1}$ and $\Gamma$, respectively, with $\Gamma ^{-1}=1.6
\, 10^{-9}s$.}
\end{figure}

\section{Evaluation of the first photon emission rate}

The emission rates can be evaluated from the matrix elements
associated with these transitions. Just after the photodissociation,
which we take as the initial time ($t=0$), the state of the two atoms
is
\begin{equation}
|\Psi _0>=N_0 (\Psi ({\bf x},{\bf y})+ \Psi ({\bf y},{\bf x}))|e>_A|e>_B
\end{equation}
with $N_0$ the normalization coefficient
\begin{equation}
N_0 =\frac{1}{(2+2Re(<\Psi ({\bf x},{\bf y})|\Psi ({\bf y},{\bf x}) >))^{1/2}}
\end{equation}
The indices $A$ and $B$ refer to the two atoms. The coordinates in
the wave function of the $A$ and $B$ atoms are denoted, respectively,
${\bf x}$ and ${\bf y}$. We represent the internal dynamics of
the atoms by the kets $|e>$ and $|g>$, which refer to the excited
and ground states. The spatial Center of Mass (CM) dynamics is
ruled by the wave function $\Psi ({\bf x},{\bf y})$. The scalar
product in the last expression represents the overlapping of the two
atoms.

The form of the initial state reflects two important physical
characteristics of the system. On the one hand, the two dissociated
atoms are identical bosons. We must symmetrize the two-boson state.
As initially they are very close, the spatial overlapping between
them is large, and exchange effects can be important. On the other
hand, the non-factorizable form of the spatial wave function
indicates that the two atoms are entangled. Here, we follow the
criterion introduced in \cite{Ghi} and \cite{Gh1} (see also \cite{ale}) to
characterize entanglement in systems of identical particles: the
state is separable if and only if $\Psi _0$ can be obtained by
symmetrization of a factorized product of two orthogonal states or
it is the product of the same state for the two particles. If $\Psi
({\bf x},{\bf y})$ is non separable, the criterion shows that $\Psi
_0$ is an entangled state. The fact that $\Psi ({\bf x},{\bf y})$ is
non-separable can be easily understood from the physical point of
view. At the classical level, the energy and momentum conservation
lead to perfect correlations between the positions and velocities of
the products of a decaying process. When we move to the quantum
realm we expect that some (nonperfect) correlations will persist. The
spatial state of the two atoms must be entangled (similarly for the
momentum representation). See \cite{Per} for a simple discussion of
the quantum correlations.

The complete two-atom and electromagnetic field initial state reads
\begin{equation}
|\Phi _0> = |\Psi _0>|0>_{EM}
\end{equation}
with the field in the vacuum state.

The first photon emission leads to the new state
\begin{equation}
|\Phi _f> = |\Psi _f>|1>_{EM}
\end{equation}
with the field changing from the vacuum to the one photon state, and the two atoms to
\begin{equation}
|\Psi _f>=\frac{1}{\sqrt 2} (\psi ({\bf x})\phi ({\bf y})|g>_A|e>_B
+ \psi ({\bf y}) \phi({\bf x})|e>_A|g>_B)
\end{equation}
This form of the atomic state is dictated by the disentanglement
process associated with the emission. It is well-known \cite{ye,ey}
that initially excited entangled states become disentangled after
spontaneous emission. This result was obtained without explicitly
taking into account the identical character of the emitters. Here, we
assume that the result remains valid when the excited atoms are
identical. For our initial state, where all the entanglement of the
system is associated with the non-separability of the wave function,
the disentanglement leads to the evolution $\Psi ({\bf x},{\bf y})
\rightarrow \psi ({\bf x}) \phi({\bf y})$. According to the
criterion in \cite{Ghi} and \cite{Gh1} state $\Psi _f$ is not entangled
(the atoms separate after the photodissociation, leading to a
negligible overlap, becoming $\psi$ and $\phi$ orthogonal).

Once we have determined the initial and first emission states, we can
evaluate the probability of this transition. The associated matrix
element is given by
\begin{equation}
{\cal M}_f =<\Phi _f |\hat{U}(t)|\Phi _0>
\end{equation}
with $\hat{U}$ the evolution operator, which can be expressed in the
form $\hat{U}=\hat{U}_{EM} \otimes \hat{U}_{at}$, with the subscript
$at$ referring to the atomic part. In addition, as the internal and
CM parts are decoupled the atomic part can be written as
$\hat{U}_{at}=\hat{U}_{CM} \otimes \hat{U}_{int}$. Moreover, as
there is no interaction between the two atoms and they are
identical (both internal operators are equal) we have
$\hat{U}_{int}=\hat{U}_{int}^{(A)} \otimes \hat{U}_{int}^{(B)}=
\hat{U}_{int}^s \otimes \hat{U}_{int}^s$, with $\hat{U}_{int}^s$ the
single-atom internal-part operator.

Using all the above expressions we can write the matrix element as
\begin{eqnarray}
{\cal M}_f = \frac{N_0}{\sqrt 2} <1|\hat{U}_{EM}|0>_{EM} <g|\hat{U}_{int}^s|e> <e|\hat{U}_{int}^s|e> \times \nonumber \\
2 (<\psi ({\bf x})\phi ({\bf y})|\hat{U}_{CM}|\Psi ({\bf x},{\bf y})>+<\psi ({\bf x})\phi ({\bf y})|\hat{U}_{CM}|\Psi ({\bf y},{\bf x})>)
\label{eq:me}
\end{eqnarray}
The coefficient $2$ in the second line of the above equation
comes from the repetition of the terms after the interchange ${\bf
x}  \leftrightarrow {\bf y}$ ($A \leftrightarrow B$).

In order to continue our calculation we note the symmetric character of the initial wave function in our arrangement:
\begin{equation}
\Psi ({\bf x},{\bf y})=\Psi ({\bf y},{\bf x})
\label{eq:sym}
\end{equation}
In effect, the Schr\" odinger equation for the system is symmetric
with respect to the variables of the two particles. This is clear
for the non-interaction terms. We also assume that the process of
dissociation leading to two bosonic atoms must be ruled by a
symmetric interaction. Thus, the initial wave function (previous to
the symmetrization associated with the bosonic character of the
particles) must be symmetric. The archetypical example of a symmetric
wave function describing the decaying process is the Gaussian one
$\Psi ({\bf x},{\bf y}) \sim \exp (-({\bf x}+{\bf y})^2/ X^2)$, with
$X$ measuring the width of the distribution. This form assures the
condition ${\bf x}_{det} \approx - {\bf y}_{det}$, representing the
quantum correlations between position detections. More elaborate
Gaussian distributions have been used in the study of dissociation.
For instance, in \cite{Fed} the authors resort to double Gaussian
functions.

Relation (\ref{eq:sym}) has two important consequences. On the one hand, the
scalar product under the normalization condition reduces to $<\Psi
({\bf x},{\bf y})| \Psi ({\bf y},{\bf x})>=<\Psi ({\bf x},{\bf
y})|\Psi ({\bf x},{\bf y})>=1$ because we assume $\Psi$ to be
normalized to unity. Then the normalization condition becomes
$N_0=1/2$. On the other hand, the two spatial matrix elements in Eq.
(\ref{eq:me}) are now equal.

The matrix element becomes ${\cal M}_f =\sqrt 2 M_s <\psi ({\bf
x})\phi ({\bf y})|\hat{U}_{CM}|\Psi ({\bf x},{\bf y})>$. In this
expression $M_s=<1|\hat{U}_{EM}|0>_{EM} <g|\hat{U}_{int}^s|e>$ is
the single particle matrix element for the excited-ground atom
transition times the vacuum-one-photon transition of the
electromagnetic field. We have used the relation
$<e|\hat{U}_{int}^s|e>=1$ for the atom that does not undergo the
transition.

From the matrix element we can derive the emission rates. In each
repetition of the experiment the emission will take place in a
random direction and we must take all these possible
directions into account. We denote by $\{\psi , \phi \}$ the set of all
possible final wave functions. ${\cal P} = <\psi |< \phi
|\hat{U}_{CM}|\Psi >$ is the probability amplitude for the evolution
$\Psi \rightarrow \psi \phi$ of the initial wave function into each
one of the members of the set. As the set is complete (the initial
wave function must necessarily evolve into one of its members) we
have $\int _{\{\psi , \phi \}} D\psi D\phi |{\cal P}|^2=1$,
representing the functional integral the sum over all the possible
final states. $|{\cal P}|^2$ and $ D\psi D\phi $ are respectively
the probability density and the volume element in the space $\{\psi
, \phi \}$. Now, we can evaluate the emission rate which reads
$\Gamma _f = \int _{\{\psi , \phi \}} D\psi D\phi  |{\cal M}_f(\psi
, \phi )|^2$, where ${\cal M}_f(\psi , \phi )$ denotes the matrix
element for each $\psi , \phi $ in the set. Note that we add
probabilities instead of amplitudes because the alternatives of
transition to $(\psi , \phi )$ and to any other pair of final wave
functions, denoted $(\psi _*, \phi _* )$, are distinguishable. We
can understand this distinguishability by noting that every pair ($\psi
, \phi $) defines a point in the functional wave-function space.
Different points in this space can be distinguished. We can reach
the same conclusion in a more technical way by comparison with the
normal space. The states associated with different points in the
normal space (in the spatial representation) are orthogonal.
Similarly, the states representing different points in the
functional space must be orthogonal and, consequently,
distinguishable. Adding probabilities we easily obtain
\begin{equation}
\Gamma _f =2|M_s|^2 \int _{\{\psi , \phi \}} D\psi D\phi |{\cal P}|^2 = 2|M_s|^2 = 2\Gamma
\end{equation}
The last step in this equation can be justified by translating the
above two-atom argument to a single atom. Introducing the set of possible
final wave functions for a single atom after spontaneous
emission and repeating step by step the argument for two excited
atoms, we have that $\Gamma =|M_s|^2$.

Two ingredients have entered into our derivation of this fundamental
relation, entanglement and identity. The initial state $\Psi _0$ is
entangled. It is simple to see that if the initial state is not
entangled the value of $\Gamma _f$ becomes different from $2\Gamma$.
We demonstrate this property in  the Appendix (Property 1).  Thus,
the initial presence of entanglement is a necessary condition to
derive Eq. (\ref{eq:ff}). The other ingredient is the identity of
the two excited atoms. The initial and final states have been
symmetrized because the atoms are identical.
This symmetrization process is fundamental to obtain the results
derived here. Without the symmetrization, that is, treating the atoms
as distinguishable, we would obtain a different value for the first
emission rate (Property 2 in the Appendix). Entanglement and bosonic
identity are necessary conditions for the derivation.

It must be remarked that the disentanglement present in the
spontaneous emission does not play a fundamental role at this stage. In order
to justify this statement we show in the Appendix (Property 3) that
Eq. (\ref{eq:ff}) is also recovered when the final state is entangled.

\section{Evaluation of the second photon emission rate}

After the first emission the state of the system is $\Phi _f$. The
atomic state after the second emission changes to
\begin{equation}
|\Psi _s>=N_s (\psi _{t_s} ({\bf x})\varphi ({\bf y}) + \psi _{t_s} ({\bf y}) \varphi({\bf x}))|g>_A|g>_B
\end{equation}
with the normalization coefficient
\begin{equation}
N_s = \frac{1}{(2+ 2|<\psi _{t_s} |\varphi >|^2)^{1/2}}
\end{equation}
where $<\psi _{t_s} |\varphi >$ is the overlap between the
two atoms at $t_s$. The subscript $t_s$ denotes the time of the second emission.

During the interval between the first and the second emissions the
wave functions evolve freely, changing from $\psi$ and $\phi$ to $\psi _{t_s}$ and $\phi _{t_s}$. After the second emission $\phi _{t_s}$ changes, due to the recoil, to $\varphi$.

The full atoms-electromagnetic-field state is
\begin{equation}
|\Phi _s>=|\Psi _s>|2>_{EM}
\end{equation}
with two photons in the field.

The second emission probability can now be evaluated from the matrix
element ${\cal M}_s =<\Phi _s|\hat{U}|\Phi _f>$, whose explicit form
is
\begin{eqnarray}
{\cal M}_s = \sqrt 2 N_s M_s (<\psi _{t_s}|\hat{U}_{CM}^{s,fr} |\psi ><\varphi |\hat{U}_{CM}^{s,re} |\phi > + \nonumber \\
<\psi _{t_s}|\hat{U}_{CM}^{s,re} |\phi ><\varphi |\hat{U}_{CM}^{s,fr} |\psi >)
\end{eqnarray}
with $\hat{U}_{CM}^s$ the single-particle CM evolution operator (the
two atoms do not interact and the two-particle operator factorizes
as $\hat{U}_{CM}=\hat{U}_{CM}^s \otimes \hat{U}_{CM}^s$). The
operators $\hat{U}_{CM}^s $ describe the generic CM evolution of the
atoms. In the case of the atom in the ground state it is a free
evolution (the wave function only spreads) and we represent it as
$\hat{U}_{CM}^{s,fr}$. For the excited atom the evolution operator
describes two steps. Between $t_f$ and $t_s$ it evolves
freely, $\hat{U}_{CM}^{s,fr}(t_f,t_s)$. Later, at $t_s$ it describes
the recoil, $\hat{U}_{CM}^{s,re}(t_s)$. We incorporate both in the
single expression $\hat{U}_{CM}^{s,re}= \hat{U}_{CM}^{s,re}(t_s)
\hat{U}_{CM}^{s,fr}(t_f,t_s)$. Using this notation we have $|\psi
_{t_s}>= \hat{U}_{CM}^{s,fr}|\psi >$ and $|\varphi >=
\hat{U}_{CM}^{s,re}|\phi >$. On the other hand, all the matrix
elements associated with the internal variables and the
electromagnetic field have been collected, as in the previous
section, in $M_s$ (now with $<2|\hat{U}_{EM}|1>_{EM}$). Combining all these expressions the matrix element reads
\begin{equation}
{\cal M}_s = \sqrt 2 N_s M_s ( 1 + <\psi _{t_s}|\varphi ><\varphi |\psi _{t_s}>)
\label{eq:vei}
\end{equation}

To evaluate the overlap between the atoms at $t_s$ we recall
that they evolve in almost-opposite directions, moving away from
each other. In short times they separate and the overlap is
small. Quantitatively, according to \cite{jap}, the separation is of
the order of $100 \mu m$, justifying the above statement. Thus, we
have $<\psi _{t_s} |\varphi > \approx 0$. Finally, we obtain
\begin{equation}
{\cal M}_s =  M_s
\end{equation}
and we arrive at $\Gamma _s = \Gamma$.

Two elements have determined this result, the disentanglement by
spontaneous emission and the negligible overlap of the two
atoms. With respect to the first element it must be stressed that if
the state after the first emission were entangled, we would
obtain a different result, as demonstrated by  Property 4 in the
Appendix. The second element is equivalent to neglecting the exchange
effects at the time of the second emission, which in our case are
represented by the second summand in Eq. (\ref{eq:vei}).

The disentanglement after the first emission is also relevant for
the experiment in \cite{jap}. After the first emission the state loses
its initial entanglement and the coincidence time spectra cannot
depend on it. We should have the same spectral distribution for any
value of the initial entanglement of the excited hydrogen atoms.
This is in marked contrast to the results in \cite{jap}, where
different spectra were obtained for different gas pressures
(identified with different entanglement values in the sample). In
the experiment in \cite{bel} the spectra were measured again for
different pressure values, obtaining in all cases the same
distribution (a decay ruled by the emission rate of single atoms)
\cite{urb}. This result completely agrees with our approach and can
be seen as a complementary verification of the disentanglement
process.

\section{Dependence on non-factorizability}

We have shown that the condition $\Gamma _f =2\Gamma $ is strongly
dependent on the nonseparable (entangled and symmetrized) character
of the initial state of the two atoms after the dissociation. In
contrast, other properties of the system such as $N_1$, $N_2$, and $N$ do
not depend on it. This behavior is a novel manifestation of a
property already described in entangled systems: nonfactorizabilty
of the state can only manifest in two-particle properties (the
values of the detection times of the two atoms are involved) of the
system. If we consider one-particle properties (we do not need to
consider both times jointly), we always obtain results compatible
with product states.

Let us examine this point in detail. When we determine the distributions
$N_1$ and $N_2$ we only care about their respective detectors. We do
not need any information about the results in the other detector. In
this case we cannot distinguish the two detection patterns from
those obtained with $2n_0$ single excited atoms (or $n_0$ pairs of
excited atoms in product states). In contrast, to determine which of
the two photons in a single repetition of the experiment is the
first one we must compare the two detection times. Now, the first
and second detection patterns are two-particle properties and differ
from that obtained for a pair of independent excited atoms.

The correlation distribution between the two emissions analyzed in
\cite{jap} is also a two-particle property as we must subtract the
times determined at the two detectors ($t_1-t_2$). However, in
contrast with the first and second photon distributions discussed
above, it does not show any dependence on the initial non-separable
character of the states. This is so because the property we measure,
$t_1-t_2$, is only related to times subsequent to the first emission.
After this emission the system is disentangled and the exchange
effects are negligible. Consequently, during that period of time the
system is in a product state. The nonseparability present in the
system for times $t<t_1$ does not affect the subsequent behavior.

\section{Discussion}

Our analysis concludes, answering the initial question, that
entanglement is actually a key ingredient in the explanation of
experiments of the type in \cite{jap} and \cite{bel}. However, it is
not the only physical ingredient involved in the problem. The
identical nature of the two atoms and the disentanglement by
spontaneous emission must also be taken into account to provide a
complete picture of the underlying dynamics. These two elements were
not considered in previous discussions. It is important to emphasize
that these ingredients enter into the picture at different stages.
Entanglement and identity act simultaneously before the first
emission and disentanglement enters the scene at the time of that
emission.

This work is a new example of the increasing interest in the role of
entanglement in atomic and molecular systems. The characterization
of entanglement is, in general, a difficult task \cite{ale}. Our
approach has the advantage of easily identifying the type of
entanglement present in the system. The form of the initial state,
$\Psi _0$, indicates that the entanglement is associated with the
spatial CM variables. Physically, this identification is a natural
consequence of the strong correlations existing between the decaying
fragments. This point of view contrasts with that expressed in
\cite{jap}, where the entanglement is between the different
constituents of the two atoms (the two electrons and the two
protons), taking into account the possible magnetic quantum numbers
of the system (see Eq. (2) in \cite{jap}). The agreement of our
results with those of the experiment in \cite{bel} shows that the
relevant variables for the problem are the CM ones and that we do
not need to be concerned about the complicated form of the connection
between the different atomic constituents.

The process of emission by excited dissociation products is an
illustrative example of entanglement and exchange effects acting at
the same time. This example suggests that this process, and decaying phenomena in general, can be a powerful tool for the
experimental study of the interplay between entanglement and
identity. The subject of entanglement in systems of identical
particles has led to many discussions and still needs clarification.
Measurement of the entanglement associated with the position and
momentum correlations of the products of decaying processes seems to
be a promising way to quantitatively study this question. In
particular, if the products of the dissociation are fermions instead
of bosons, we should expect different patterns. This extension to
fermions could be an interesting line of research.

Finally, and this is the most important result from the practical
point of view, we have shown that setups of the type in \cite{jap} and \cite{bel} provide an experimental verification of the process of
disentanglement by spontaneous emission. As we indicated earlier it is,
to the best of our knowledge, the first demonstration of the effect. This
result widens the scope of the field of experimental disentanglement
by including the case associated with spontaneous emission. It also
opens the doors to explore other aspects of the field. For instance,
one can study and test whether the process depends on the type of
particles involved (distinguishable, bosonic or fermionic) or on the
initial degree of entanglement.

\section*{Appendix}

In this Appendix we demonstrate some properties used in the text.

{\bf Property 1.} According to the criterion in \cite{Ghi} and \cite{Gh1} the form of a nonentangled initial state must be
\begin{equation}
|\Psi _0^{NE}>=N_0^{NE} (\chi ({\bf x})\xi ({\bf y}) + \chi
({\bf y}) \xi({\bf x}))|e>_A|e>_B
\end{equation}
with $N_0^{NE} =(2+2|<\chi |\xi >|^2)^{-1/2}$, denoting by $<\chi
|\xi >$ the overlap of the two one-particle wave functions. In addition,
$\chi$ must be either orthogonal to $\xi$ ($N_0^{NE}=1/\sqrt 2$) or
equal to it ($N_0^{NE}=1/2$).

Repeating the steps used to derive ${\cal M}_f$ we obtain
\begin{eqnarray}
{\cal M}_f^{NE}= \sqrt 2 N_0^{NE} M_s \times \nonumber \\
(<\psi |\hat{U}_{CM}^s|\chi><\phi |\hat{U}_{CM}^s|\xi> + <\psi
|\hat{U}_{CM}^s|\xi> <\phi |\hat{U}_{CM}^s|\chi>)
\end{eqnarray}
with $\hat{U}_{CM}^s$ the single-particle CM evolution operator
introduced in Sec. IV. When $\chi$ is orthogonal to $\xi$, in the
probabilities derived from the amplitude ${\cal M}_f^{NE}$ we have
interference terms between $<\psi |\hat{U}_{CM}^s|\chi><\phi
|\hat{U}_{CM}^s|\xi>$ and $<\psi |\hat{U}_{CM}^s|\xi> <\phi
|\hat{U}_{CM}^s|\chi>$. In general this amplitude probability does
not lead to the relation $\Gamma _f =2 \Gamma $. Only when $\chi
=\varphi $ do we obtain the correct value for $\Gamma _f$. However, the
condition of equality of the two one-particle wave functions leads
to the relations ${\bf x}_{det} \approx {\bf y}_{det}$, which shows
that this particular case is not relevant for our dissociation
problem.

{\bf Property 2.} If we do not symmetrize the initial state it reads
$|\Psi _0^{NS}>= \Psi ({\bf x},{\bf y})|e>_a |e>_b$. Note that we
use lowercase letters $a$ and $b$ to label the atoms instead of
capital ones in order to show that now it is possible in some way
to distinguish the particles (if the particles were
indistinguishable it would be necessary to symmetrize). After the
first emission the state can be $|\Psi _f^{NS}>=\psi ({\bf x}) \phi
({\bf y})|g>_a |e>_b $ or $|\tilde{\Psi } _f^{NS}>=\tilde{\psi}
({\bf x}) \tilde{\phi }({\bf y})|e>_a |g>_b $ with equal
probabilities $1/2$. The probability amplitude for the first channel
is ${\cal M}^{NS}=<\Psi _f^{NS}|\hat{U}|\Psi _0>$, and the
expression for the second channel is similar. To obtain the transition probability we must sum both probabilities with equal weights $1/2$: as we have
assumed that the two atoms are somehow distinguishable, it is in
principle possible to distinguish which one has emitted the photon
and both emission alternatives become distinguishable; we must add
probabilities instead of probability amplitudes. Now, using the same
arguments as in the text, we have ${\cal M}^{NS}=\tilde{\cal
M}^{NS}=M_s$ and, finally, $\Gamma ^{NS}=\Gamma$.

{\bf Property 3.} When the final state is entangled we have
\begin{equation}
|\tilde{\Psi }_f>=\frac{1}{\sqrt 2} (\tilde{\Psi }({\bf x},{\bf y})|g>_A|e>_B + \tilde{\Psi }({\bf y},{\bf x})|e>_A|g>_B)
\end{equation}
Repeating the steps in Sec. III and, in particular, invoking the symmetry condition of $\Psi$ we obtain
\begin{equation}
\tilde{{\cal M}}_f =\sqrt 2 M_s <\tilde{\Psi }({\bf x},{\bf y}) | \hat{U}_{CM}|\Psi ({\bf x},{\bf y})>
\end{equation}
Performing a summation over the continuous set of possible final states $\tilde{\Psi}$ we recover Eq. (\ref{eq:ff}).

{\bf Property 4.} We assume that the state after the first emission
remains entangled and, consequently, is given by $\tilde{\Psi }_f$.
If there is not disentanglement after the first emission, we can
safely assume that there is also not disentanglement after the second
one and  the atomic state will remain entangled,
\begin{equation}
|\tilde{\Psi }_s>= \tilde{N}_s (\Psi _*({\bf x},{\bf y}) + \Psi _*({\bf y},{\bf x})) |g>_1|g>_2
\end{equation}
with $\tilde{N}_s=(2+2Re(<\Psi _* ({\bf x},{\bf y})|\Psi _*({\bf y},{\bf x}) >))^{-1/2}$.

The transition matrix element in this case is
\begin{eqnarray}
<\tilde{\Psi}_s|\hat{U}|\tilde{\Psi}_f> = \sqrt 2 \tilde{N}_s M_s \times \nonumber \\
(<\Psi _*({\bf x},{\bf y})|\hat{U}_{CM}|\tilde{\Psi}({\bf x},{\bf y})> + <\Psi _*({\bf x},{\bf y})|\hat{U}_{CM}|\tilde{\Psi }({\bf y},{\bf x})>)
\end{eqnarray}
When evaluating the probabilities from the matrix element we find
interference terms between the spatial matrix elements in the above
equation. Moreover, there is a dependence, through the normalization
factor, on the final overlap. These two results prevent a
relation of the type $|<\tilde{\Psi}_s |\hat{U}|\tilde{\Psi}_f>|^2 =
\Gamma$. Another form of reaching the same conclusion is to consider
the particular case of both wave functions, $\tilde{\Psi}$ and $\Psi _*$,
symmetric. For this case we would obtain $|<\tilde{\Psi}_s
|\hat{U}|\tilde{\Psi}_f>|^2 = 2\Gamma \neq \Gamma$.


\begin{thebibliography}{99}

\bibitem{jap} T. Tanabe, T. Odagiri, M. Nakano, Y. Kumagai, I. H. Suzuki, M. Kitajima, and N. Kouchi, Phys. Rev. A {\bf 82}, 040101(R) (2010).
\bibitem{com} P. Sancho and L. Plaja, Phys. Rev. A {\bf 83}, 066101 (2011).
\bibitem{bel} X. Urbain, A. Dochain, C. Lauzin, and B. Fabret, J. Phys. Conf. Ser. {\bf 635} 112085 (2015).
\bibitem{ye} T. Yu and J. H. Eberly, Phys. Rev. Lett. {\bf 93} 140404 (2004).
\bibitem{ey} T. Yu and J. H. Eberly, Science {\bf 323} 598 (2009).
\bibitem{bra} M P Almeida, F de Melo, M Hor-Meyll, A Salles, S P Walborn, P H Souto Ribeiro and L. Davidovich, Science {\bf 316}, 579 (2007).
\bibitem{urb} X. Urbain private communication. Results contained in an extended version of \cite{bel} to be published.
\bibitem{Ghi} G. C. Ghirardi, L. Marinatto, and T. Weber, J. Stat. Phys. {\bf 108}, 49 (2002).
\bibitem{Gh1} G. C. Ghirardi and L. Marinatto, Phys. Rev. A {\bf 70}, 012109 (2004).
\bibitem{ale} M. C. Tichy, F. Mintert, and A. Buchleitner, J. Phys. B {\bf 44}, 192001 (2011).
\bibitem{Per} A. Peres, Am. J. Physics {\bf 68}, 991 (2000).
\bibitem{Fed} M. V. Fedorov, A. E. Kazakov, K. W. Chan, C. K. Law, and J. H. Eberly, Phys. Rev. A {\bf 69}, 052117 (2004).
\end{thebibliography}
\end{document}